\documentclass[prd,twocolumn,showpacs,floatfix,amsmath,amssymb,floatfix,nofootinbib]{revtex4}
\usepackage{graphicx,color,dcolumn,booktabs,bm}
\usepackage{longtable,lscape,comment,braket}
\usepackage{txfonts}
\usepackage{overpic}
\usepackage{amssymb}
\usepackage{indentfirst}
\usepackage{feynmf}
\usepackage{slashed}
\usepackage{cases}
\usepackage{epstopdf}
\usepackage{psfrag}
\usepackage{subfigure}
\pdfoutput=1
\usepackage{color}
\usepackage{multirow}

\graphicspath{{Figures/}} %
\usepackage[colorlinks, citecolor=blue,anchorcolor=red,menucolor=red, linkcolor=red,filecolor=red,runcolor=red,urlcolor=blue,frenchlinks=red]{hyperref}

\begin{document}

\title{Study of structures and dynamical decay mechanisms for multiquark systems}
\author{Xuewen Liu$^{1}$}
\email{liuxuewen@mail.nankai.edu.cn}
\author{Hong-Wei Ke$^2$}
\email{khw020056@tju.edu.cn}

\author{Xiang Liu$^{3,4}$}
\email{xiangliu@lzu.edu.cn}

\author{Xue-Qian Li$^{1}$}
\email{lixq@nankai.edu.cn}

\affiliation{$^1$School of Physics, Nankai University, Tianjin 300071, China \\
             $^2$School of Science, Tianjin University, Tianjin 300072, China\\
             $^3$School of Physical Science and Technology, Lanzhou University, Lanzhou 730000, China\\
             $^4$Research Center for Hadron and CSR Physics, Lanzhou University \& Institute of Modern Physics of CAS, Lanzhou 730000, China}

\begin{abstract}
The inner structures of the multiquark states are an interesting subject in hadron physics, generally they may be in  tetraquark states which are composed of colored constituents, or in molecular states which are composed of two color singlets, or their mixtures. Therefore, the mechanisms which bind the constituents in a unique system and induce the multiquark states to decay would be different in those cases. In this work, using the quantum mechanics we analyze the dynamical mechanisms inducing decays of the tetraquarks
where $Y(4630)$ stands as an example for the study, we also comment on the molecular states without making numerical computations.

\end{abstract}
\maketitle

\section{Introduction}
\label{sec:intro}
To study the inner structures of multiquark states is an interesting subject in hadron physics, because it
is obviously beyond the simple $q\bar q$ or $qqq$ textile which was established by Gell-Mann more than half century ago.
The molecular structures of the multiquark systems have been studied by many authors, in comparison the tetraquark states have much less
been investigated so far. Generally, the tetraquark is suggested to be composed of a diquark and an antidiquark which reside in
color antitriplet and triplet, respectively~\cite{Maiani:2004vq, Brodsky:2014xia,Montanet:1980te}. Because of the different color structures,
the dynamics for molecular states and tetraquark would be very different. In fact, in molecular states, the two constituents are bound by exchanging
color-singlet mesons or baryons which can be described by the chiral Lagrangian. Instead, the two constituents in tetraquarks are bound by direct
gluon exchange. As for the quarkonium, the interaction between quark and antiquark is realized by exchanging
gluons, a single-gluon exchange composes the leading contribution which results in a Coulomb-type effective potential, however, for the hadron
formation, the energy scale is below $\Lambda_{QCD}$, therefore the nonperturbative QCD effects would play an important role which induces
a confinement piece in the potential. Similarly, for the tetraquark case, the diquark and antidiquark interact by exchanging gluons and definitely the
single gluon exchange exists and induce a Coulomb-like potential whereas the nonperturbative QCD effects should also be introduced. It is generally
believed that such interaction may be described by the color-flux model~\cite{Brodsky:2014xia}. Following this scenario, we will study the dynamics
which not only results in the different structures of the two configurations, but also determines their different decay patterns. Because the newly observed
$Y(4630)$ is very likely to be a tetraquark, it would be an ideal place for carrying research on the teraquark and molecular states via their decay behaviors.

$Y(4630)$ has been observed in the invariant mass spectra of the $e^+e^-\to\Lambda_c\bar\Lambda_c$ channel~\cite{Pakhlova:2008vn}, and
it is identified as a $J^{PC}=1^{--}$ resonance with mass $M=4634^{+9}_{-11}$ MeV and width $\Gamma=92^{+41}_{-32}$ MeV.

There are many alternative interpretations for the observed peak~\cite{Chen:2016qju,Esposito:2014rxa}, for example,
the authors of Ref.~\cite{Lee:2011rka} consider that a strong attraction
between $\Lambda_c$ and $\bar\Lambda_c$ bind them together, so that $Y(4630)$ may be interpreted as a  baryon-antibaryon molecule.
Instead, in Refs.~\cite{Badalian:2008dv,Segovia:2008ta}  $Y(4630)$ is interpreted as a $5^3S_1$ charmonium state.
Also, the $Y(4630)$ is considered to be induced by a threshold effect  instead of
being a genuine resonance~\cite{vanBeveren:2008rt}.

Among those proposals, the suggestion that $Y(4630)$ is a tetraquark state is more favorable \cite{Cotugno:2009ys,Maiani:2014aja,Liu:2016sip}.
In Ref.~\cite{Maiani:2014aja}, $Y(4630)$ is identified as the ground state with its orbital angular momentum $L=1$.
By reanalyzing the $\Lambda_c\bar\Lambda_c$ and $\psi(2S)\pi^+\pi^-$ spectra, Cotugno
\textit{et al.}  suggested that  $Y(4630)$ and
$Y(4660)$~\cite{Wang:2007ea,Wang:2014hta} could be the same tetraquark state, and is the first radial excitation of the
$Y(4360)$ with $L=1$~\cite{Cotugno:2009ys}.
In Ref.~\cite{Liu:2016sip}, the authors studied the open-charm decay $Y(4630)\to \Lambda_c\bar \Lambda_c$ by assuming that $Y(4630)$ is a
radially
excited state of the diquark-antidiquark bound state with hidden charm.
By another theoretical physics group \cite{Guo:2010tk} $Y(4630)$ is  interpreted as a molecular state made of $\psi(2S)$ and $f_0(980)$.

As is well known, in general the multiquark states may be in tetraquark states which are composed of colored constituents, or in molecular states which are composed of two color singlets, or their mixtures. Therefore, the mechanisms which bind the components in a unique system and induce the multiquark states to decay would be different in those cases. In this work, using the quantum mechanics we analyze the dynamical mechanisms inducing decays of the tetraquarks and the molecular states, where $Y(4630)$ stands as an example for the study. Thus,
a numerical analysis on the decay width of channel $Y(4630)\to\Lambda_c\bar{\Lambda}_c$ based on the tetraquark postulate is made and a qualitative discussion about
the possible decay process $Y(4630)\to\psi(2S)\pi^+\pi^-$ where $Y(4630)$ may be a mixture of tetraquark and molecular state is presented.

The paper is organized as follows: after this introduction, we study the decay of $Y(4630)$ with the tetraquark  and molecular interpretations in Secs~\ref{sec:tetra} and~\ref{sec:mole} respectively, then Sec~\ref{sec:conclusion} is devoted to our discussion and conclusion.

\section{Tetraquark picture}
\label{sec:tetra}
Inspired by the fact that the $Y(4630)$ decays into charmed baryon
pair, one is tempted to conjecture this resonance as a tetraquark which is made of the
diquark-antidiquark  $[cq][\bar c\bar q]$, where $q$ is a light quark either $u$ or $d$,  $[cq]$ resides in a color antitriplet whereas $[\bar c\bar q]$ is in a color triplet.

Here we take the diquark-antidiquark picture proposed by Brodsky \textit{et al.} \cite{Brodsky:2014xia} that  a diquark and an antidiquark are
bound together by a gluon-flux-tube into a color singlet tetraquark where the constituents are separated by a substantial distance once they are
created. The interaction for the system can be well described by  a generalized Cornell potential~\cite{Eichten:1978tg} since the constituents (diquark and antidiquark) are treated as two pointlike color sources in analog to the configuration for an ordinary $Q\bar Q$ quarkonium.

In terms of the quark pair creation (QPC) model where a quark-antiquark pair is excited out from vacuum, we calculated the decay width of $Y(4630)\to \Lambda_c\bar\Lambda_c$~\cite{Liu:2016sip}. In that picture, the QCD vacuum is excited and a
quark-antiquark pair is created. The quark-antiquark pair would ``tear" apart the diquark and antidiquark due to strong QCD interaction between quark and diquark (antiquark
and antidiquark). Then joining the diquark, the created quark becomes a constituent of the charmed baryon $\Lambda_c$, and as well for the $\bar\Lambda_c$.

In parallel, let us consider an alternative way to discuss the production of $\Lambda_c\bar\Lambda_c$ in the framework of quantum mechanics.

In fact, the color flux-tube results in a potential barrier to forbid the inner constituents (either the diquark or the antidiquark) to escape from the bound state.
In terms of a modified flux-tube-induced potential,
the decay of $Y(4630)$ may occur as the diquark-antidiquark bound system falling apart via tunneling through the effective potential barrier.
After tunneling out the barrier, the diquark (antidiquark) would immediately attract a quark (antiquark) from the vacuum to compose a color singlet
$\Lambda_c$ ($\bar\Lambda_c$) and the hadronization process is somehow similar to the picture frequently used to
study the multiparticle production at high energy collision.
Then the transition probability can be calculated in terms of the WKB (Wentzel-Kramers-Brillouin) approximation.

\subsection{Potential model}\label{potential}
First, we employ a nonrelativistic potential model with a  Cornell-like potential where some free parameters are obtained by
fitting the mass spectrum of heavy quarkonia and generalized to the case for tetraquark, then we get the wave function of $Y(4630)$
by solving the Schr\"odinger equation.

The general Hamiltonian of a diquark-antidiquark system (i.e. a quarkonium-like system)  can be written as
\begin{equation}\label{eq:Ham}
 H=\frac{{\bf p}^2_1}{2m_1}+m_1+\frac{{\bf p}_2^2}{2m_2}+m_2+V(r) ,\\
\end{equation}
where the $m_1({\bf p}_1)$ and $m_2({\bf p}_2)$ are the masses(3-momenta) of the diquark $cq$ and antidiquark $\bar c\bar{q}$ respectively, we take $m_1=m_2=m$ in this work. The interaction potential is
\begin{equation}\label{eq:V}
V(r)=V_{\text{oge}}(r)+V_{\text{conf}}(r),
\end{equation}
where $r$ is the distance between the diquark and the antidiquark. The one-gluon-exchange (oge) term $V_{\rm {oge}}(r)$, which plays the main role at short distances,  is \cite{Barnes:2005pb}
\begin{equation}\label{eq:Voge}
V_{\text{oge}}(r)=-\frac{4}{3}\frac{\alpha_s}{r} + \frac{32\pi\alpha_s}{9m^2}{\mbox{\boldmath $S$}}_1\cdot{\mbox{\boldmath $S$}}_2 \delta (r),
\\
\end{equation}
and the confinement part $V_{\text{conf}}(r)$ takes the linear form\cite{Cornell}
\begin{equation}\label{eq:Vcon}
V_{\text{conf}}(r)=b r+c,
\end{equation}
where ${\mbox{\boldmath$S$}}_{1(2)}$ and $-4/3$ are, respectively, the spin operators and the color factor specific to  ${\mbox{\boldmath$3$}}$-${\mbox{\boldmath$\bar 3$}}$ attraction.
$b$ is the string tension and $c$ is a global zero-point energy.  $\alpha_s$ is the phenomenological strong coupling constant.

The $\delta$-function in Eq.(\ref{eq:Voge}) is replaced by a Gaussian smearing function \cite{Weinstein:1983gd} with a fitted parameter $\sigma$
\begin{equation}\label{eq:delta}
\delta(r)\rightarrow \Big(\frac{\sigma}{\sqrt{\pi}}\Big)^3e^{-\sigma^2r^2}. \\
\end{equation}

The spin wave functions for the $Y(J^{PC}=1^{--})$ state with $L=1$ is taken as  $Y_1 = |0,0,0,1\rangle_1$ in the basis of $|S_{qc},S_{\bar q \bar c},S_{\rm total},L\rangle_{J=1}$ ~\cite{Maiani:2014aja}.

Here we take the values from Ref.~\cite{Barnes:2005pb}: $\alpha_s=0.5461, b=0.1425$ GeV$^2, c=0, \sigma=1.0946$ GeV.

By adopting the suggestion given by authors of Refs.~\cite{Cotugno:2009ys,Liu:2016sip} that  $Y(4630)$
is a radial excitation state of P-wave (L=1), the (anti)diquark mass is determined to be $1.878$ GeV
which is close to the $D$ meson mass. This value is also consistent with that computed by using QCD sum rules in Ref.~\cite{Kleiv:2013dta}
where it is $1.86\pm0.05$ GeV.

\begin{center}
\begin{figure}[htbp]
\includegraphics[width=5.6cm]{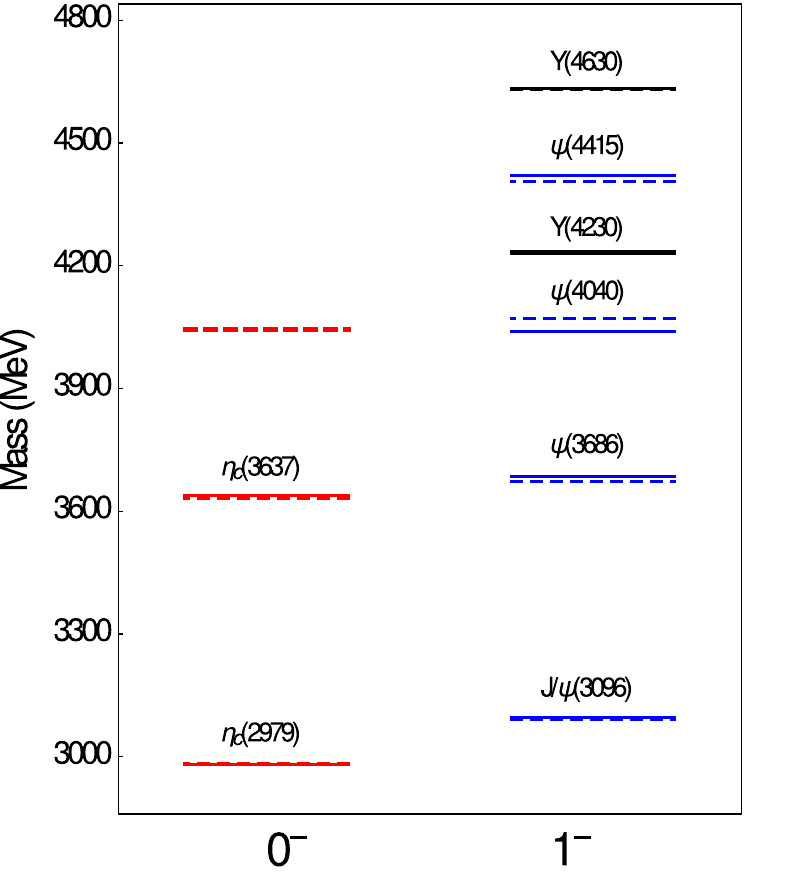}
\caption{The fitted spectra. We also present the  charmonium spectra obtained in Ref.~\cite{Barnes:2005pb}. The dashed lines stand for the computed masses, and the solid ones are experimental values taken from the data book~\cite{Agashe:2014kda}.}
\label{Fig:spectra}
\end{figure}
\end{center}

\begin{center}
\begin{figure}[htbp]
\includegraphics[width=6.5cm]{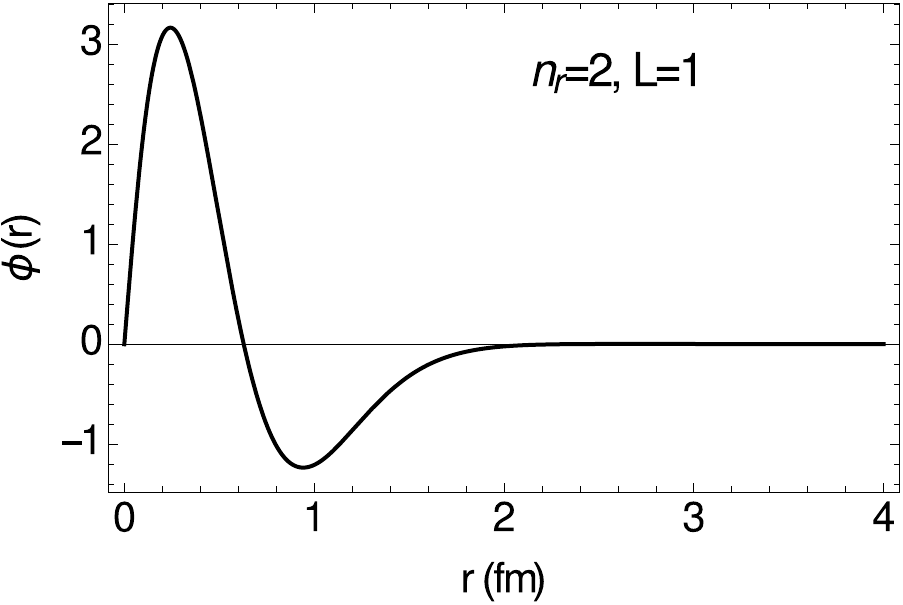}
\caption{The radial wave function $\phi(r)$ of $Y(4630)$ with quantum numbers $n_r=2, L=1$. }
\label{Fig:wave}
\end{figure}
\end{center}

The fitted spectra are presented in Fig.~\ref{Fig:spectra}, and the charmonium spectra calculated by authors of Ref.~\cite{Barnes:2005pb} are
also shown in the figure for a clear comparison. In this framework, we fit the ground state to be 4235 MeV, and such a
state is consistent with the observed $Y(4230)$ resonance~\cite{Ablikim:2014qwy}  and/or $Y(4220)$ resonance~\cite{Chang-Zheng:2014haa,Yuan:2014rta}
which is also considered as a tetraquark ~\cite{Faccini:2014pma}.
The radial wave function of $Y(4630)$ is plotted in Fig.~\ref{Fig:wave}.

\subsection{Decay of  $Y(4630)$  as a tetraquark}\label{WKB}

The interaction between the constituents in the tetraquark cannot simply be derived from the quantum field theory yet and phenomenologically
the dynamics of the nonperturbative QCD effects which determines the confinement behavior may be described by the flux-tube model.
Moreover, as is well known, when the tension of the flux-tube goes beyond a certain bound, namely, the distance between the diquark and
antidiquark gets long enough, the flux-tube will break into two strings and at the new ends a quark-antiquark pair
is created~\cite{Kokoski:1985is,Kumano:1988ga}. One can use
a step function to describe the breaking effect as
\begin{equation}
 [1-\theta(r-r_0)]\times V(r),
\end{equation}
where $r_0$ is a parameter corresponding to the strengthening limit of the string at where the probability of the string fragmentation reaches maximum. A typical scale for nonperturbative QCD is $\Lambda_{{\rm QCD}}$, therefore
it is natural to consider $r_0$ should be of order of $ \sim 1/\Lambda_{{\rm QCD}}$. Just as smearing the delta function, we need also to smear the step function. In fact
$$1-\theta(r-r_0)=\lim_{\epsilon\to 0}\frac{1}{\text{e}^{\frac{(r-r_0)}{\epsilon r_0}}+1},$$
so smearing the step function implies that we keep $\epsilon$ as a nonzero free parameter to be determined.
In fact, if we do not consider breaking of the flux tube, the effective potential is the same as $V(r)$ given in Eq.~(\ref{eq:V}). Therefore,
taking into account of the ``breaking" effect does not affect the computation on the $Y(4630)$ spectrum.

Here the interaction between the diquark and the antidiquark at a relatively large distance is described by a modified potential as~\cite{Liu:2014dla}.
\begin{equation}\label{eq:flux-con}
V'(r)=V(r)\frac{1}{\text{e}^{\frac{(r-r_0)}{\epsilon r_0}}+1}.
\end{equation}

\begin{center}
\begin{figure}[htbp]
\includegraphics[width=7.5 cm]{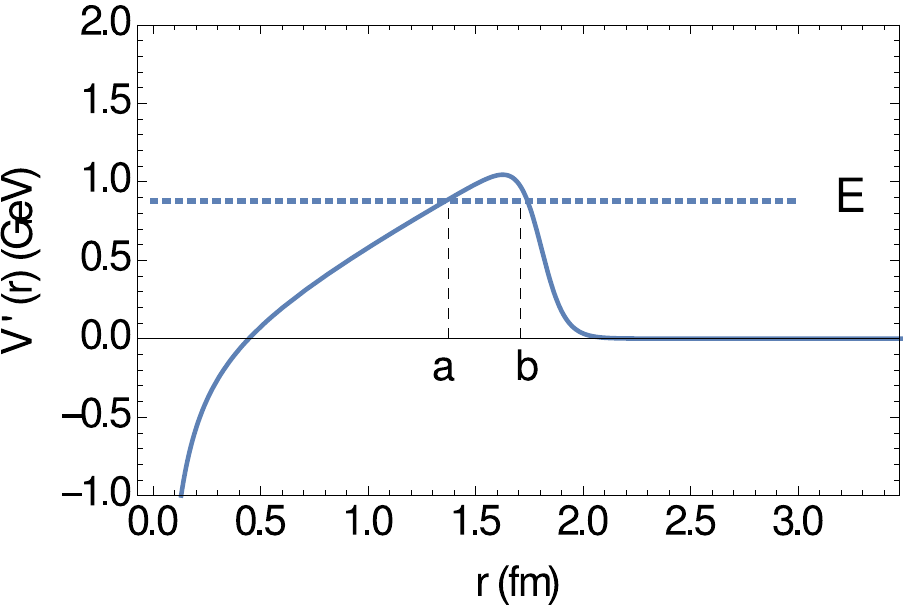} 
\caption{The decay mechanism for $Y(4630)$. The dotted line stands for the eigenvalue $E$, and $a, b$ are the turning points at the potential barrier.
In this figure, we set $\epsilon$ and $r_0$ to be 0.03 and 1.8 fm respectively for an illustration.}
\label{Fig:tunnel}
\end{figure}
\end{center}

In this scenario, we translate the flux-tube induced confinement into the potential barrier, and  breaking the tube corresponds to tunneling
through the barrier.
The diquark-antidiquark bound system falls apart by tunneling through this effective potential barrier, then is hadronized into a  $\Lambda_c \bar{\Lambda}_c$ pair.
The process is graphically shown in Fig.~\ref{Fig:tunnel}.
By means of the WKB approximation, the transition probability of the tunneling process can be calculated.

Under this assignment, the transition probability is given by
\begin{equation}\label{eq:traniation}
T=\text{exp}[-\frac{2}{\hbar}\int_a^b\sqrt{2\mu(E-V'(r))}~dr],
\end{equation}
where $\mu=m_1m_2/(m_1+m_2)=m/2$ is the reduced mass of the $[cq]$-$[\bar{c}\bar{q}]$ system, $a$ and $b$ are the turning points, as shown in Fig.~\ref{Fig:tunnel}.

One can obtain the effective velocity $v$ of the motion of a particle with the reduced mass inside the system  from the average
kinetic energy $\langle\psi({\bf r})|\frac{{\bf p}^2}{2m}|\psi({\bf r})\rangle$  where $\psi({\bf r})$ is the wave function obtained in terms of the
pure Cornell potential without the
flux tube breaking correction. Since the breaking effect only affects the long distance behavior, it does not change the wave function $\psi(r)$ and
the decay width $\Gamma_Y=\frac{v}{2a} T$ is deduced.

\begin{center}
\begin{figure}[htbp]
\includegraphics[width=7.5 cm]{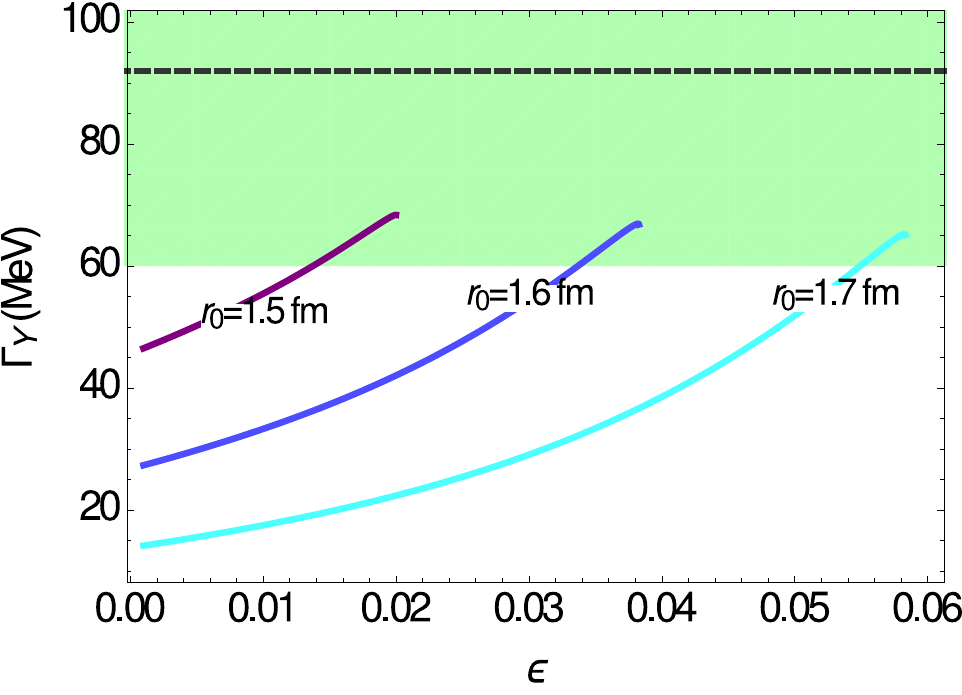}
\caption{\label{results} Dependence of the partial decay width of $Y(4630)\to\Lambda_c\bar{\Lambda}_c$ on $\epsilon$,
where $Y(4630)$ is supposed to be in $n_r=2$ state, the
Belle data are shown in the plot for a comparison.  The dashed black line and the green
band correspond to the central value and error for the total width of
$Y(4630)$ measured by the Belle collaboration ($\Gamma=92^{+41}_{-32}$ MeV~\cite{Pakhlova:2008vn}).
The purple, blue and cyan curves correspond to the three different $r_0$ assignments respectively.}\label{h1}
\end{figure}
\end{center}

Assuming the $Y(4630)$ as the first radial excited state,
we compute the partial decay width of the channel $Y(4630)\to\Lambda_c\bar\Lambda_c$.
In Fig.~\ref{results}, dependence of the calculated partial width $\Gamma_Y$ on the free parameter $\epsilon$ is plotted.
The purple, blue and cyan curves correspond to the cases of $r_0=1.5, 1.6, 1.7$ fm, respectively. As $\epsilon$ increases, the predicted decay width increases
as shown  in Fig. \ref{h1}, where each curve in the figure corresponds to a special $r_0$ value, and it is noted that each curve stops at some value
of $\epsilon$, because at that $\epsilon$ value, the diquark and antidiquark are no longer bound by the modified potential.
It sets a constraint on  $\epsilon$. The dashed black line and the green band are respectively the central value and the error of the total width of
$Y(4630)$ measured by the Belle collaboration ($\Gamma=92^{+41}_{-32}$ MeV)~\cite{Pakhlova:2008vn}.
Varying $r_0$  which represents the flux-tube breaking effect does not change the general picture.

It is noticed that the calculated  partial decay width changes with $r_0$, but for any specific value of $r_0$, there exists
a range of $\epsilon$  which allows the predicted width to be consistent
with the experimental data, namely, falls within the error tolerance region set by present measurements, which
is rather wide.  
Indeed, even though the calculated decay width of $Y(4630)\to\Lambda_c\bar{\Lambda}_c$ is smaller than the
central value of the measured total width, one still cannot conclude that $Y(4630)$ is not a pure tetraquark yet.

Our numerical results indicate that the tetraquark picture is able to predict the correct decay width
of $Y(4630)$, even though not completely confident,
we believe that its decay mechanism could be considered as the diquark-antidiquark system falling apart
via tunneling through an effective potential barrier, and then diquark and antidiquark are respectively hadronized into color singlet hadrons, and
the $\Lambda_c\bar\Lambda_c$ pair should be the main product.
This conclusion is consistent with that of Ref.~\cite{Liu:2016sip}, in which a similar result is obtained within the QPC framework.

It is noteworthy that if the future
measurement indeed confirms a rather large total width which is larger than our prediction
based on the pure tetraquark structure, a possible mixture between tetraquark and molecular state should be taken into account
and other decay modes such as $Y(4630)\to\psi(2S)\pi^+\pi^-$ may occur with non-negligible fraction.

\section{Molecular picture}
\label{sec:mole}

As is mentioned in the introduction and the above discussion, the hadronic molecular picture $\psi' f_0(980)$ is another possible choice for the inner structure of $Y(4630)$
which was proposed by Guo {\it et al.}~\cite{Guo:2010tk}. It is noted that the mass and width of $Y(4630)$ are consistent
with those for the $Y(4660)$ state ($M=4652\pm 10\pm8$ MeV, $\Gamma=68\pm 11\pm1$ MeV~\cite{Wang:2014hta}) within error tolerance.
By taking into account the $\Lambda_c \bar{\Lambda}_c $ final state interaction, it is found that the $Y(4630)$ may be the same state as $Y (4660)$,
and the resonance can be a $\psi' f_0(980)$ molecular bound state.

\begin{center}
\begin{figure}[htbp]
\includegraphics[width=7.5cm]{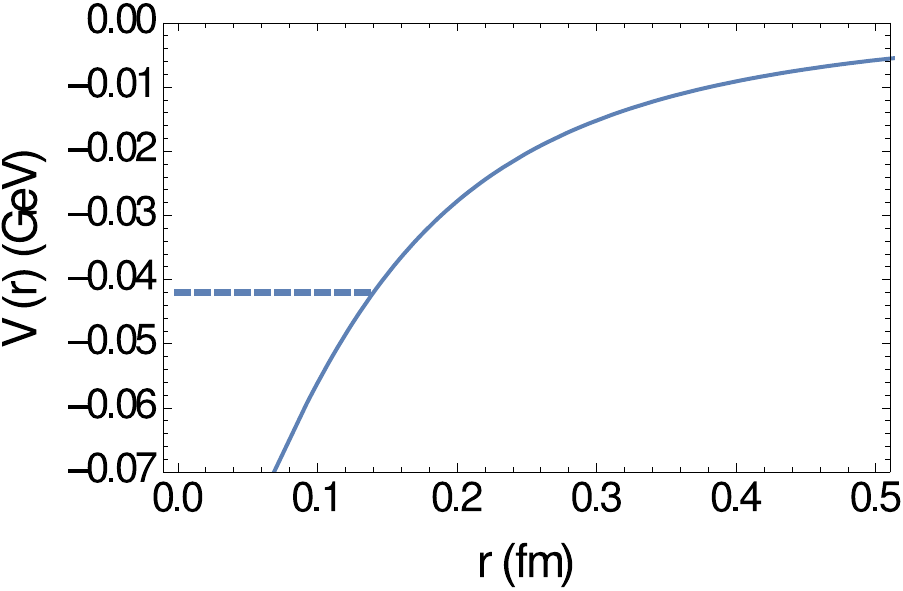}
\caption{Diagram illustrating the constituents $\psi(2S)$ and $f_0(980)$ of $Y(4630)$ residing in the attractive potential well which is supposed to be
a hadronic molecular state.  The effective potential is induced by the $\sigma$ meson exchange, see Ref.~\cite{Liu:2008tn} for details.} \label{Fig:well}
\end{figure}
\end{center}

Now let us study the mechanism which may induce the decay of the molecular state.
Because the two constituents in the molecular state are color singlets, they do not interact by directly exchanging gluons, but only via
exchanging color-singlet mesons, and the leading contribution is coming from the  $\sigma(f_0(600))$ exchange between $\psi(2S)$ and $f_0(980)$
which does not induce a potential barrier as for the tetraquark case. Instead, the interaction provides a potential well and the constituents are confined in
the well, as shown in Fig.~\ref{Fig:well}.
Thus we should have $M_{Y(4630)}=M_{\psi(2S)}+M_{f_0(980)}+\Delta E$ where $\Delta E$ is the binding energy of the molecular state and roughly is $-30$ MeV.
In the traditional framework of quantum mechanics it is a stable structure, i.e. $Y(4630)$ cannot dissolve
into on-shell $\psi(2S)$ and $f_0(980)$,  however, due to the quantum fluctuation, $f_0(980)$ can jump out the potential well to become an off-shell virtual
particle. If the virtual $f_0(980)$ does not transit into two real pions, it would fall back to its original state inside the molecule.
The duration of it being virtual particle can
be estimated by the uncertainty principle as $\Delta \tau\cdot |\Delta E|\sim {\hbar\over 2}={1\over 2}$ in the natural  unit system, thus
the virtuality time $\Delta\tau$ is proportional to ${1\over |\Delta E|}$ where $\Delta E$ is the binding energy. Obviously the decay amplitude should be proportional
to $\Delta\tau$, namely the larger the binding energy $|\Delta E|$ is, the shorter  $\Delta \tau$ is, and then the smaller the decay probability would be.
By this principle, we can write out an effective Lagrangian which induces the decay of the molecule $Y(4630)\to \psi(2S)+f_0^*(980)\to \psi(2S)+\pi\pi$ where the superscript
$*$ denotes that $f_0(980)$ is an off-shell virtual meson which later transits into two pions. The effective vertex at $Y(4630)-\psi(2S)f_0(980)$ is
\begin{equation}
L={g\over |\Delta E |}A^{\dagger}_{\mu}\overleftrightarrow\partial_{\alpha}B^{\mu}\partial^{\alpha}\phi,
\end{equation}
where $A_{\mu},\; B_{\mu}$ and $\phi$ correspond to $\psi(2S),\; Y(4630)$, $f_0(980)$ and $g$ is a dimensionless universal coupling constant. By the equation of
motion it is easy to be reduced into $L'$ which reads as
\begin{equation}
L'={g(m_B^2-m_A^2)\over |\Delta E|}A^{\dagger}_{\mu}B^{\mu}\phi.
\end{equation}
The effective coupling given by authors of Ref.~\cite{Guo:2010tk}  is $\frac{g'^2}{4\pi}=4(M_{\psi'}+m_{f_0(980)})^2\sqrt{2|\Delta E|/\mu'}$ where $\mu'$ is the reduced mass of the $\psi'$ and $f_0(980)$,  $g'$ has an energy dimension.
Thus we can identify the relation
\begin{equation}
{g(m_B^2-m_A^2)\over |\Delta E|}=g'.
\end{equation}

\begin{center}
\begin{figure}[htbp]
\scalebox{0.25}{\includegraphics{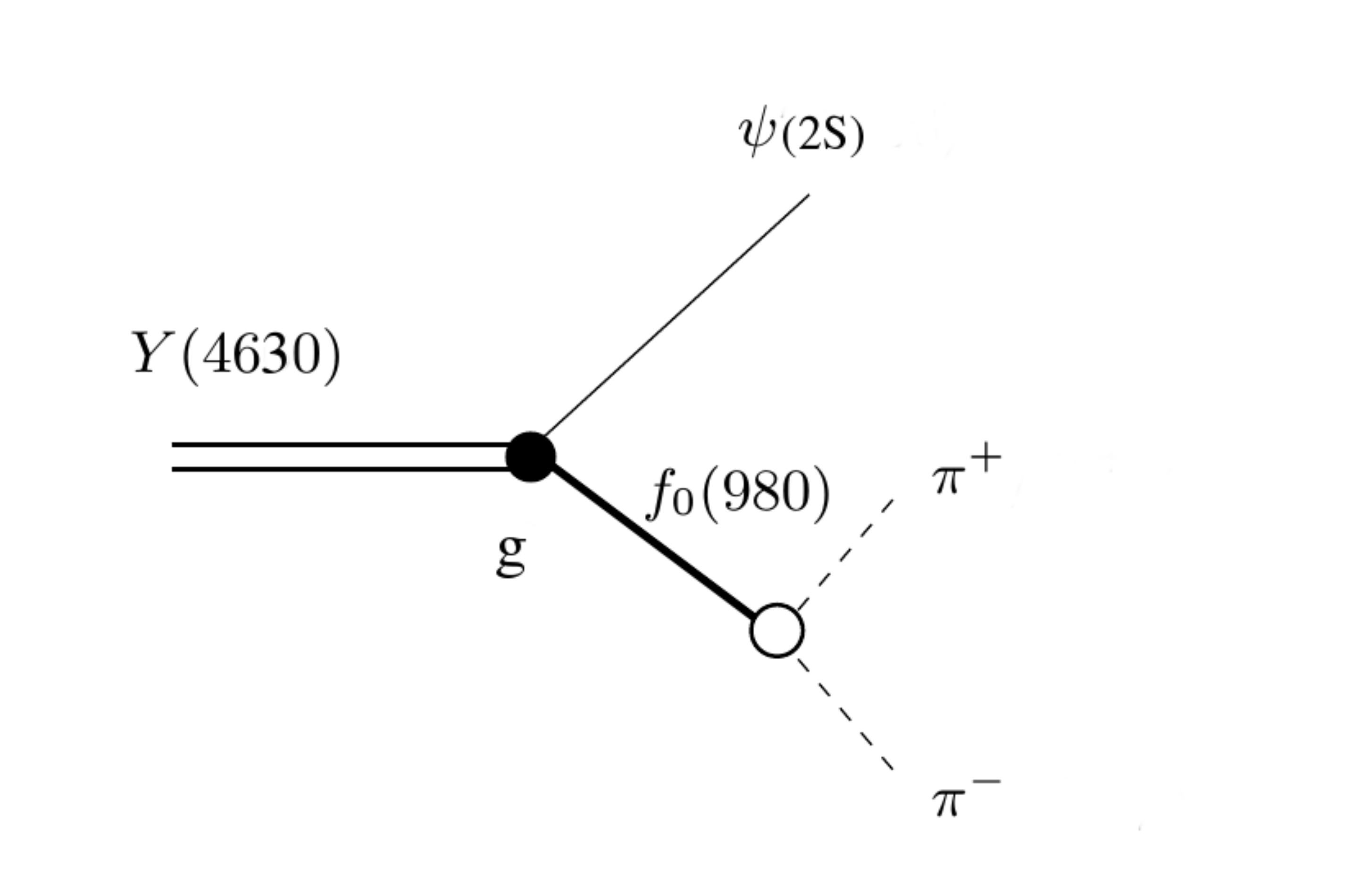}}
\caption{Diagram illustrating a possible decay channel of the $Y(4630)$ in the molecular picture which is $Y(4630)\to\psi(2S)\pi^+\pi^-$.} \label{otherdecay}
\end{figure}
\end{center}

With this assumption another decay mode of $Y(4630)$ could be $Y(4630)\to \psi(2S)\pi^+\pi^-$ for the off-shell $f_0(980)$ mainly decaying into $\pi^+\pi^-$, as shown in Fig.~\ref{otherdecay}.

With the given Lagrangian, the decay width was calculated in Ref.~\cite{Guo:2010tk} as $\Gamma(Y\to\psi(2S)\pi^+\pi^-) = 8$ MeV, here we do not repeat it and
advise readers to refer to that paper.

\section{Conclusion and discussion}
\label{sec:conclusion}

The hadronic decay is closely associated with the nonperturbative QCD, and a lot of phenomenological models are proposed to
account for its effects. For example,
the QPC model, flux-tube model, QCD sum rules and
lattice QCD,  etc. have been successfully used to estimate decay rates, even though, with the
exception of the lattice calculation, none of them can be directly derived from quantum field theory so far.

For the $Q\bar Q$ systems, the physics picture is clear, even though a phenomenological model must be embedded
to reflect the nonperturbative
QCD effects and the computation schemes are mature. However, for the four-quark states, the
inner structure and dynamics which leads the binding and decay of the state are still not well understood and there are various proposals for them. In this work we study the decay mechanisms of $Y(4630)$ in both tetraquark and molecule pictures in the framework of quantum
mechanics. Namely, we use the  WKB approximation to calculate the decay width of $Y(4630)$ as it is assumed to be a tetraquark state, and then qualitatively
discuss its decay mechanism as it is postulated as a molecular state where $f_0(980)$ jumps out the potential well due to a quantum fluctuation and becomes
a virtual particle and later transits into two real pions.

Definitely, all of the assignments to the observed resonance at 4630 MeV should be tested in the future by more precise measurements. In our other works~\cite{Liu:2016sip,Guo:2016iej}, we study the case that if $Y(4630)$ is a tetraquark, its favorable decay mode should be $Y(4630)\to\Lambda_c\bar\Lambda_c$ which would
overwhelmingly dominate its width, but due to the inelastic rescattering processes between $\Lambda_c$ and $\bar\Lambda_c$, some other final states, such as
$p\bar p$, $n\bar n$, $D^{(*)}\bar D^{(*)}$ and $\pi\pi,\; K\bar K$ might be produced with measurable rates, whereas, if $Y(4630)$ is a molecular state, its
dominant decay mode would be $\psi(2S)\pi\pi$ and due to the decay of $\psi(2S)$ and final state interaction, the pattern of the decay products which will
be experimentally measured would be completely different from the tetraquark case. Thus the measurements would provide more information about the assignment
of $Y(4630)$. We are lying hope on the future experiments which will be carried out at the BELLEII, BESIII and even LHCb in the coming years.

Moreover,  we suspect if there is a mixing between the tetraquark  and molecular states which results in $Y(4630)$ and $Y(4660)$, it would be an interesting picture.

\vspace{2mm}
\acknowledgments{We sincerely thank Prof. Mu-Lin Yan for helpful discussion on the issue of quantum fluctuation. This work is supported by the National Natural Science Foundation of China under the contract No. 11375128, No. 11135009, No. 11222547 and No. 11175073. Xiang Liu is also supported by the National Youth Top-notch Talent Support Program (``Thousands-of-Talents Scheme").}


\end{document}